# Social Networking: an astronomer's field guide

E. L. Gomez, H. L. Gomez, J. Yardley

1. Las Cumbres Observatory,6740 Cortona Dr. Suite 102, Santa Barbara, CA 93117, USA.  2. School of Physics and Astronomy, Cardiff University, Queens Buildings, The Parade, Cardiff CF24 3AA, UK.  3. Faulkes Telescope Project, Cardiff University, Queens Buildings, The Parade, Cardiff CF24 3AA, UK

**Summary**. **We present a brief introduction to the phenomenon of "social networking" and its potentially powerful use as an astronomy outreach and educational tool. We briefly discuss the development of applications for websites and facebook and the use of web trackers e.g. Google Analytics to analyze your audience. Finally we discuss how social bookmarking can be used to promote your work to unexpected audiences.**

## 1 Introduction

Social networking is simply a way of keeping in touch with other users and sharing multimedia web content with friends and acquaintances via the internet.  Over the past 5 years, social networking has becoming increasingly popular. Previous incarnations consisted of users subscribing to an email list, where the user would need to actively select a topic of their interest (often highly specialized) and then passively sit back while all the traffic from the email list filled their inbox, usually never to be read again. There is still a place for this type of network structure, but to be truly effective in a social context and as users become more technologically sophisticated, a more interactive approach is necessary to fully engage users with a real social experience.

Recently, social networking has seen a huge increase in popularity thought to be due to the novel sharing aspect; it provides people with a way to share interests with a self-selected group (commonly called 'friends', irrespective of whether they are relatives, colleagues or casual acquaintances). Users connect with others based on relationships or common ideals.  Most social networking sites do not discriminate and are used by millions; consequently they can be an extremely powerful medium for educational purposes.  In this article we discuss the power of social networking and bookmarking for astronomy outreach.

## 2 Social Networks

According to the internet surveying organization, comScore, the internet has 890 million users worldwide, 550 million of which are members of a social networking site; 62% of the total global users are clustered around a relatively low number of these social network website, mainly facebook, Bebo and MySpace. By taking advantage of these pre-existing media you can massively increase the traffic to your website or educational tool. Thus, if you want to promote your astronomy or education project more widely, this is a good place to look for a large audience.

To promote your information/site effectively, it helps to know a little about the system. The propagation of information across your social network can be limited by many factors. Mainly, the number of friends you have and the likelihood that they will pass  information on to non-mutual friends. For example, you have an educational video you wish to promote so you send it (or share) to all your friends. If you only have 5 friends and they are not regular users, your venture will not meet with much success.

Although the sites are non-discriminating, not all represent everyone equally; the median age of social networking sites is 21 but many users are much younger particularly in the popular MySpace site.  Class divisions are also obvious in different social networks with facebook users tending to come from white middle class backgrounds, often under 35 and are people who value higher education.  An analogy is that of newspaper readership; some people look for current affairs, others prefer gossip columns, and many flip straight to the sports pages.  Knowing your audience is very impor-tant, and once you have identified your target audience, it is important to choose the relevant network(s). One advantage to the sites is that users can select their groups according to the simple criteria of whether a particular person is known to them (and usually whether they like them or not), and not because of a highly specialized interest (as with email groups and forums), which allows educators/scientists to reach out to larger communities.

## 3 Social Bookmarking

Social bookmarks also allow users to share interests and materials but in some senses the "network" part of social bookmarking is the exact opposite to that utilized by sites like facebook and MySpace. A social bookmarker will nominate categories which they are interest in (e.g. astronomy, politics, $17^{th}$ century costume etc.) and the websites 'delivered' to them will reflect these choices. The delivery mechanism requires the bookmarker to visit the bookmarking website; Digg, StumbleUpon, Del.icio.us, and Reddit are some examples. Ways to nominate content to be bookmarked come in two main varieties: via a web browser toolbar or via links directly on your webpage.

Many social bookmarking sites also provide a small application plugin which can be installed into your favorite web browser. This allows you to nominate any webpage to be bookmarked, and this website is then promoted on the bookmarking website for other users to see. The more often users bookmark a particular site, the higher its page rank becomes and the more prominently the bookmark site will promote it, and the more users will see it.

Some examples of the bookmarking websites available are:
- Digg - provides just the number of distinct times a site has been bookmarked. If people do not like a site but still bookmark it, the site will rise to the top, regardless of opinions on the website itself.
- Del.icio.us – this was originally founded to provide a way of allowing web users to have an online record of their personal bookmarks. The website is very simple and uncluttered, unlike many others of its kind. It also allows each user to add their own tags to a bookmarked site.

- StumbleUpon – this site stands apart from the rest because it allows its users give either a 'thumbs up' or 'thumbs down' rating to a website.

By posting a bookmark link on your website you can more actively solicit your visitors to bookmark the site. This has the advantage that you are already directing people to nominate your site and you have editorial control over factors such as how the page is categorized.  These websites also allow the bookmarker to write comments or a miniature review of the site. StumbeUpon can provide you with immediate feedback from a potentially huge pool of users on the effectiveness and look of your site and if your project is deemed popular, the World is your oyster. One caveat: if everyone bookmarks everything, this system does not work.

## 4  Developing Applications for Facebook

Facebook is, at the time of writing, the most popular social networking website with over 140 million users (with 5 million different users regularly active each month). This site stands apart from the rest by allowing the community to write their own applications to share with friends and other users. Thus there are now over 100,000 facebook developers with applications advertising their project. The community contributed applications range wildly from simply asking the user questions e.g. to determine which Doctor Who character they resemble, to the more advanced e.g. games where users can play their friends.

Making an application is relatively straightforward to anyone who has programming experience in web languages (e.g. php, html, css). Learning facebook's bespoke XML markup, fbML, is not too taxing and you can find some pleasant surprises waiting for you there. For example, where you would normally think of having to make a database call to translate a user's unique ID number to their name, there is an fbML tag which does this for you. The facebook developers' guide is not as comprehensive as it could be but will provide you with some basic steps to follow and also provides a sample application to cut your teeth on.

An application has two main aspects; a box and a page. The box aspect can be placed on your profile page (your homepage within the site). The page aspect is the homepage of your application (see **Fig.1**). In addition to these there is also a developer homepage to which you can add photos and videos, use discussion boards and members can become fans of your application; all of the standard facebook features that user profiles have.

The potential to applications for keeping your audience involved and feeling a part of a project is worth the time spent on developing the applications. There are, however some limitations. There is no facility to dynamically refresh the content you provide in a box (the component which is placed on the user's profile page) by running a cron job on the facebook server. The content in the box is served from a static file, but the page content is loaded dynamically. Therefore when a user accesses your application's page, you can dynamically generate new content for the box, but not by just viewing the box on your profile. If your application provides content which only changes once per day, it might be sufficient, to reply on your userbase visiting the application's page often (which, in practice, happens rarely).

One solution to this is to set a cron job to run on a non-facebook server (either one you have control over or a web cron service, e.g. http://www.webcron.org/). There is a further trick which involves setting a, so-called, infinite session key to gracefully side-step the need to be logged into facebook to access the page.

One other complication arises when the application tailors the box content to a particular user. In this case when you refresh the content, you must also find each user's settings and perform the update individually. This might seem a fairly trivial thing to do and is arguably the sort of thing computers are good at. However if your application has 1000 users and updates every 10 minutes, there is a lot of wasted data transferring between the sites. Even the keenest users will not spend more than a few hours on the site per day (typically statistics show the average time is one hour) and they won't always be staring at their profiles waiting for your application to update.

An example of a facebook application is shown in Fig. 1 (written by E. Gomez) where the latest astronomical image taken with telescopes owned by the Las Cumbres Observatory Global Telescope network (LCOGT) is shown, and refreshes every 5 minutes. The latest observation box is the dynamic content which is refreshed whenever a user views this page. This box is the content displayed in the user's profile box. Viewing the page view updates facebook's cache of the box. This application, although simple, allows the general public to feel part of the LCOGT team and is also an effective way to keep scientists up-to-date with the latest images and news.

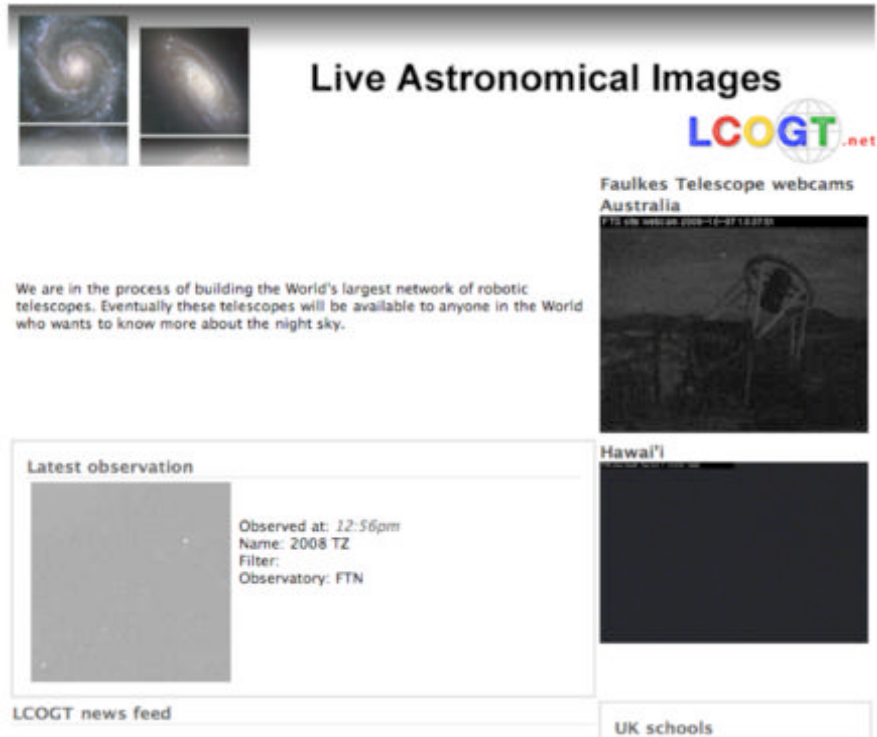

**Fig. 1.** Screenshot of the LCOGT "latest images" Facebook application page. This application is designed to update fans and users with the latest images from the two 2-m LCOGT telescopes.

## 5 Case Study: The Impact Calculator

As part of the Down2Earth educational program run by the National Museum of Wales, we (E. Gomez and J. Yardley) developed a new interactive online Astronomy tool to demonstrate the science behind Earth impacts and the creation of impact craters. The Impact Calculator (available at http://www.down2earth.eu/impact_calculator) allows users to simulate smashing an asteroid into Earth and see how big a crater their asteroid made. The idea was based on an academic version used originally as a research tool by J. Marcus, H. Melosh and G. Collins based on the work in Collins et al. (2005). Our approach here was to create an online tool designed to be visually appealing to UK school children in particular, but also the largest audience possible.

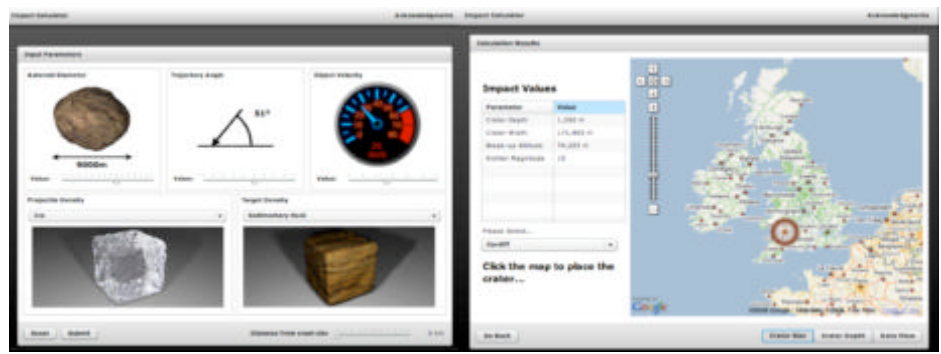

**Fig. 2.** Screenshots from the Impact Calculator. The user can vary the parameters of the incoming projectile. After choosing the projectile parameters, the user selects *submit* and they are directed to a Google Map to centre their crater (red circle).

With this tool, users can choose the projectile velocity, diameter and composition (ice, porous or dense rock and iron) as well as the composition of the ground (water, sedimentary rock and igneous rock) and trajectory angle. An example is provided in **Fig. 2**. The Calculator allows users to smash an asteroid into the Earth at various locations: London, Cardiff, Paris, New York and the Barringer Crater site in Arizona using Google Maps. The crater created from the user-input parameters is placed over Cardiff in this example. (The size of the impact crater created by the projectile is shown by the red circle which indicates the extent of the crater.) Users can also compare the depth of their impact crater with well-known landmarks such as the Eiffel Tower and Big Ben as well as investigating some of the effects of the impact e.g. whether or not the projectile has broken before reaching the ground or if the user has succeeded in creating total mass destruction. The frequency of such an collision event and the impact energy in Joules are also provided in the *Data View* section allowing students to study how the different im-pact parameters chosen for the asteroid change the impact event and the after effects. The Impact Calculator is available in six different languages (English, Welsh, Spanish, French, German and Polish).

The Impact Calculator is an interesting example to highlight the power of social networking and social bookmarks. After its release online, it was quickly picked up by the internet community and bookmarked on many of the sites metioned above. The Calculator has not received any promotion, news articles or press release since it has not been officially launched. The site is also not officially linked to from any other source and the results we present here are solely from web traffic as a result of social bookmarking and blogs.

The number of individual hits the Impact Calculator has received to date is over 100,000 (in the period 1 July to 1 Oct 2008). The geographical distribution of the web traffic is shown in **Fig. 3** (the map and statistics provided by Google Analytics). This result was surprising, especially as the site had not been publicized, yet it received hits from web users across the globe simply due to its ranking in bookmarking sites.

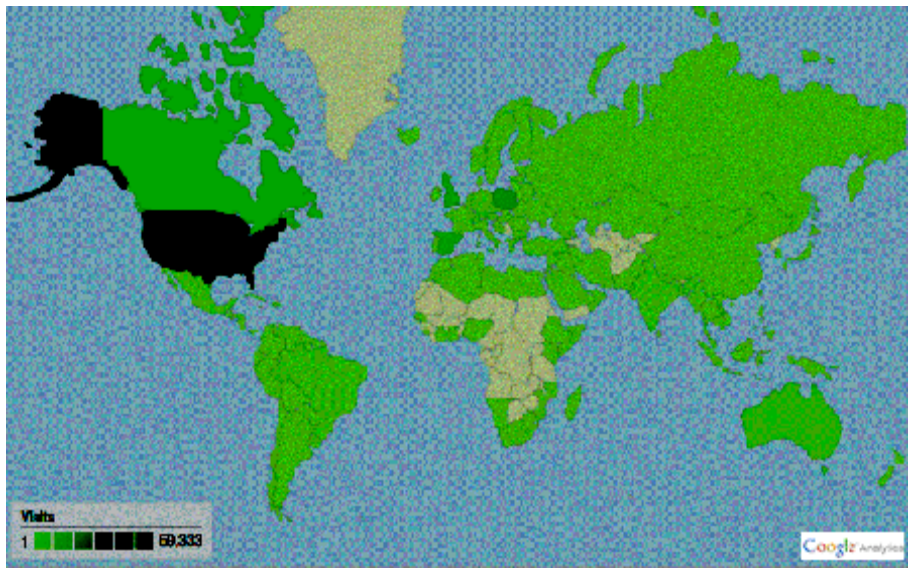

**Fig. 3.** World map showing locality of users obtained from Google Analytics. The colour scale represents the number of users accessing the Impact Calculator from July – October 2008.

The majority of the traffic (56,000 users) was provided by the social bookmarking site, StumbleUpon (**Fig. 4**). The traffic from the USA is significantly higher than any other country and accounts for approximately half of the total visits. The next highest hits are from users in Poland (after being publicized on their popular Astronomy site: Astronomia.pl), leaving UK in fourth position. Although the application was originally created to specifically support an education programme in Wales, with the Calculator being the first Welsh-language online astronomy tool, it has received less than 100 hits from this region. Although the number of users is pleasing, annoyingly, we do not appear to have reached our target audience!

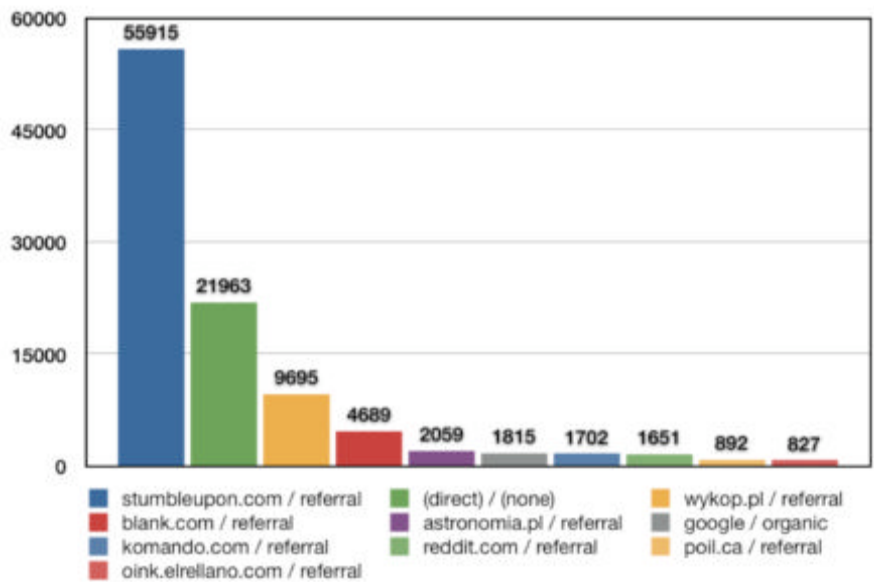

**Fig. 4.** The frequency of users referred to the Impact Calculator site from various other sites. The effects of social bookmarking is clearly seen here with most traffic arriving from StumbleUpon.

## 6 Summary

Making your applications public while they are in beta testing or before the final features have been enabled, can lead to valuable (and often unexpected) user comments and suggestions. Make sure there is a way people can contact you (even if it an anonymous address such as info@mysite.com).

To really understand your audience, install Google Analytics or similar web statistic program , the results can tell you far more than just the number of hits and allow you to determine the reach of an outreach/educational/scientific project, ideal for future grant applications.

The social network experience can be addictive. If you have a message to promote, these networks can help you spread it to a huge audience. Make your site is attractive and update it regularly to give people a reason to keep visiting it.  Read user comments on blogs or bookmarking sites and act on them, this is particularly valuable if you want to reach the general public, it really does make users feel part of the process.

## Acknowledgements

*The Impact Calculator was developed as part of Down2Earth, an educational project funded by an STFC Science Centre Award, held by the National Museum of Wales.*